\documentclass{elsart5p}
\usepackage{graphics}
\usepackage{graphicx}
\usepackage{epsfig}
\usepackage{amssymb}
\usepackage{amsmath}

\newcommand{\Ayy}{\mbox{$A_{y,y}$}}
\newcommand{\Axx}{\mbox{$A_{x,x}$}}
\newcommand{\Azz}{\mbox{$A_{z,z}$}}
\newcommand{\vpvpdpi}{\mbox{$\pol{p}\pol{p}\to d\pi^+$}}
\newcommand{\ppdpi}{\mbox{$pp\to d\pi^+$}}
\newcommand{\vnvpdpi}{\mbox{$\pol{n}\pol{p}\to d\pi^0$}}
\newcommand{\npdpi}{\mbox{$np\to d\pi^0$}}

\begin{document}
\begin{frontmatter}

\title{First measurements of spin correlations in the $\pol{n}\pol{p}\to d\pi^0$ reaction}
\author[ikp,dubna]{V.~Shmakova},
\author[ikp,tbilisi]{D.~Mchedlishvili},
\author[ikp,dubna]{S.~Dymov\corauthref{cor1}},
\ead{s.dymov@fz-juelich.de}
\author[dubna]{T.~Azaryan},
\author[gatchina]{S.~Barsov},
\author[ikp,tbilisi]{D.~Chiladze},
\author[ikp]{R.~Engels},
\author[ikp]{R.~Gebel},
\author[Lanzhou,ikp]{B.~Gou},
\author[ikp,gatchina]{K.~Grigoryev},
\author[ikp]{M.~Hartmann},
\author[ikp]{A.~Kacharava},
\author[dubna]{V.~Komarov},
\author[cracow]{P.~Kulessa},
\author[dubna]{A.~Kulikov},
\author[dubna]{V.~Kurbatov},
\author[tbilisi]{N.~Lomidze},
\author[ikp]{B.~Lorentz},
\author[dubna,tbilisi]{G.~Macharashvili},
\author[ikp]{S.~Merzliakov},
\author[ikp,gatchina]{M.~Mikirtytchiants},
\author[ikp,gatchina]{S.~Mikirtytchiants},
\author[tbilisi]{M.~Nioradze},
\author[ikp]{H.~Ohm},
\author[ikp]{D.~Prasuhn},
\author[ikp]{F.~Rathmann},
\author[ikp,dubna]{V.~Serdyuk},
\author[ikp]{H.~Seyfarth},
\author[ikp]{H.~Str\"oher},
\author[tbilisi]{M.~Tabidze},
\author[rossendorf,MSU]{S.~Trusov},
\author[dubna]{D.~Tsirkov},
\author[dubna]{Yu.~Uzikov},
\author[ikp,gatchina]{Yu.~Valdau},
\author[ucl]{C.~Wilkin}
\corauth[cor1]{Corresponding author.}

\address[ikp]{Institut f\"ur Kernphysik and J\"ulich Centre for Hadron
Physics, Forschungszentrum J\"ulich, D-52425 J\"ulich, Germany}
\address[dubna]{Laboratory of Nuclear Problems, Joint Institute for Nuclear
  Research, RU-141980 Dubna, Russia}
\address[tbilisi]{High Energy Physics Institute, Tbilisi State University, GE-0186
Tbilisi, Georgia}
\address[gatchina]{St. Petersburg Nuclear Physics Institute, RU-188350 Gatchina,
  Russia}
\address[Lanzhou]{Institute of Modern Physics, Chinese Academy of Sciences, Lanzhou 730000, China}
\address[cracow]{Institute of Nuclear Physics, PL-31342 Cracow,
Poland}
\address[rossendorf]{Institut f\"ur Kern- und Hadronenphysik,
Forschungszentrum Rossendorf, D-01314 Dresden, Germany}
\address[MSU]{Skobeltsyn Institute of Nuclear Physics, Lomonosov Moscow State University, RU-119991 Moscow, Russia}
\address[ucl]{Physics and Astronomy Department, UCL, Gower Street, London WC1E 6BT, United Kingdom}


\date{Received: \today / Revised version:}

\begin{abstract}%
The transverse spin correlations \Axx\ and \Ayy\ in the \vnvpdpi\ reaction
have been measured for the first time in quasi-free kinematics at the
COSY-ANKE facility using a polarised deuteron beam incident on a polarised
hydrogen cell target. The results obtained for neutron energies close to
353~MeV and 600~MeV are in good agreement with the partial wave analysis of
data on the isospin-related \ppdpi\ reaction, though the present results
cover also the small-angle region, which was largely absent from these data.
\end{abstract}

\end{frontmatter}

It follows from isospin invariance that the cross section for \npdpi\ should
be half of that for \ppdpi\ but all the spin observables should be identical
for the two reactions. However, there have been relatively few measurements
of the neutron-induced cross section~\cite{HUT1990,OPE2003} and even less is
known about the spin dependence.

In contrast, the \ppdpi\ reaction has long provided a test bed for countless
phenomenological models of pion production at intermediate energies, some of
which are summarised in Refs.~\cite{GAR1990,HAN2004}. In more recent years
the reaction has also been frequently discussed within the framework of
chiral perturbation theory~\cite{FIL2012}. The available data set is very
large, with measurements of the cross section, analysing powers, spin
correlations and spin transfers in both the direct and inverse channel and
these have allowed partial wave analyses to be attempted for proton beam
energies $T_p < 1.3$~GeV~\cite{ARN1993}.

There have been several measurements of the longitudinal $A_{z,z}$ and the
transverse spin correlation \Ayy\ in the \vpvpdpi\ reaction, sometimes
leading to inconsistent results~\cite{ARN1993}, but meaningful data on \Axx\
are much rarer~\cite{APR1984}. If one keeps only the dominant $S$-wave
$\Delta(1232)N$ intermediate state then $\Axx=\Ayy=\Azz=-1$ in the forward
direction and so it is not surprising that the measured values of these
observables are strongly negative at intermediate energies. The combination
$\Axx+\Ayy -\Azz+1$ must vanish when the pion c.m.\ polar angle
$\theta_{\pi}=0^{\circ}$ or $90^{\circ}$ and deviations from this at
arbitrary angles only arise from pion $d$ or higher waves. As a consequence,
quite generally \Axx\ can never be positive in the forward
direction~\cite{WIL1980}.

In a free two-body or quasi-two-body reaction, such as \npdpi\ or
$np\to\{pp\}_{\!s}\pi^-$, we take the beam direction to lie along the
$z$-axis and the momentum of the produced pion in the c.m.\ to be
$\vec{p}_{\pi}=
(p_{\pi}^x,p_{\pi}^y,p_{\pi}^z)=p_{\pi}(\sin\theta_{\pi}\cos\phi_{\pi},
\sin\theta_{\pi}\sin\phi_{\pi},\cos\theta_{\pi})$. If the target $Q$ and beam
$P$ polarisations are in the $y$ direction then the dependence of the
differential cross section on the pion azimuthal angle $\varphi_{\pi}$ is
given by
\begin{eqnarray} \nonumber
&&\frac{d\sigma}{d\Omega}=\left(\frac{d\sigma}{d\Omega}\right)_{\!0}
\left[1 + (PA_y^P + QA_y^Q)\cos\varphi_{\pi}\right.\\
&&\hspace{2.55cm}\left.+PQ\left(\Ayy\cos^2\varphi_{\pi} +
\Axx\sin^2\varphi_{\pi}\right)\right].
\label{asymmetry}
\end{eqnarray}
The unpolarised c.m.\ differential cross section $(d\sigma/d\Omega)_0$ and
the beam $A_y^P$ and target $A_y^Q$ analysing powers, as well as the
spin-correlation parameters \Axx\ and \Ayy, are all functions of
$\theta_{\pi}$.

However, for a quasi-free reaction induced by an incident deuteron, the
neutron direction is not precisely aligned along that of the beam due to the
Fermi motion in the deuteron. This introduces an incident angle that can be
several degrees in the laboratory system. In order to correct for this
effect, the three-momentum of the incident neutron is reconstructed on an
event-by-event basis. Nevertheless, any small $P_z$ and $Q_z$ components
arising in such a quasi-free measurement are neglected.

A second effect of the Fermi motion, especially its longitudinal component,
is that in a quasi-free reaction the c.m.\ energy $\sqrt{s}$ is not
unambiguously fixed by the incident beam energy. Although in our case the
effective beam energy $T_{\rm neutron} = [s-(m_p+m_n)^2]/2m_p$ is constructed
for every event, the limited statistics and the finite resolution mean that
data must be grouped over a relatively wide energy range.

Two experiments were carried out at the COoler SYnchrotron (COSY) at the
Forschungszentrum J\"ulich using the ANKE magnetic
spectrometer~\cite{BAR1997}, which is sited at an internal target position of
the accelerator. Though the purposes of the measurements were different, the
apparatus and analysis techniques were quite similar. The first used
transversally polarised 726~MeV deuterons to investigate the spin-correlation
in the $\pol{n}\pol{p} \to \{pp\}_{\!s}\pi^-$ reaction in the vicinity of
$T_{\rm neutron} \sim 353$~MeV, where the diproton $\{pp\}_{\!s}$ is
dominantly in the $^{\!1}S_{0}$ state~\cite{DYM2013}. The second used
1,200~MeV deuterons to study the $\pol{d}\pol{p}\to \{pp\}_{\!s}n$ reaction
at around 600~MeV per nucleon, as a precursor for measuring $np$
charge-exchange amplitudes at higher energies~\cite{MCH2013}.

The apparatus was described in great detail in the two
publications~\cite{DYM2013,MCH2013} and only the salient points are discussed
here. The spectrometer consists of two dipole magnets D1 and D3, which
deflect the circulating deuteron beam from and back to the nominal orbit,
respectively. The main dipole D2 is used to analyse the momenta of charged
particles produced in a nuclear reaction. Only the ANKE Forward Detector
(FD)~\cite{CHI2002,DYM2003} was needed to study the \npdpi\ reaction in both
of these experiments.

A measurement of a spin correlation requires both a polarised beam and
target. Although a polarised target could be provided directly in the form of
a jet from the polarised Atomic Beam Source (ABS)~\cite{MIK2013}, much higher
densities can be achieved if this is used to feed a storage cell. Cells of
dimensions $x\times y\times z = 19\times 15\times 390$~mm$^3$ (at 726~MeV)
and $20 \times 15 \times 370$~mm$^3$ (at 1,200~MeV) were constructed from
25~$\mu$m Teflon-coated aluminium. Here $z$ is measured along the beam
direction with $x$ and $y$ referring to the horizontal and vertical
transverse directions, respectively.

Having registered two fast charged particles in the forward detector,
deuteron-proton pairs were isolated by comparing the difference between their
arrival times in the FD with that predicted from the reconstructed
three-momenta of the particles. At 726~MeV this was confirmed by the energy
loss in the forward hodoscope. In order to enhance the quasi-free nature of
the reaction, only events where the kinetic energy of the detected proton in
the rest frame of the incident deuteron was below 6~MeV were retained. In
this case it is expected that this is a \textit{spectator} proton, $p_{\rm
spec}$, that only influences the reaction through the kinematics. The
distributions in the proton momenta, which are shown explicitly in
Refs.~\cite{DYM2013,MCH2013}, are consistent with this interpretation.

\begin{figure}[h]
\begin{center}
\includegraphics[width=1.0\columnwidth, angle=0]{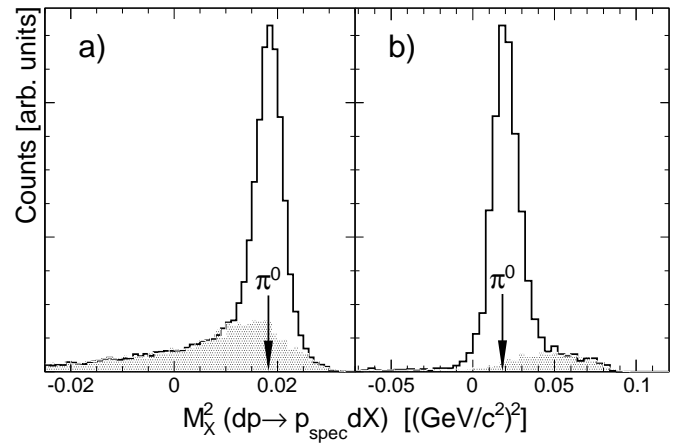}
\caption{\label{mm} Comparison of the $(d,dp_{\rm spec})$
missing-mass-squared distributions when using a polarised
hydrogen target (solid lines) or filling the cell with nitrogen
gas (shaded area). a) $T_d=726$~MeV, b) $T_d=1,200$~MeV.}
\end{center}
\end{figure}

After putting a cut on the spectator momentum, the $dp\to dp_{\rm spec}\pi^0$
reaction was finally identified from the missing-mass peak in the $dp\to
dp_{\rm spec}X$ data, which is illustrated for the two energies in
Fig.~\ref{mm}. The vertex position is clearly less well determined when using
a long cell rather than a point target and there can also be contamination
arising from the beam halo striking the cell walls. The shape of this
missing-mass background was simulated by filling the cell with nitrogen gas.
This was then fitted outside the peak region and subtracted to reveal the
$\pi^0$ signal.

\begin{figure}[h]
\begin{center}
\includegraphics[width=1.0\columnwidth, angle=0]{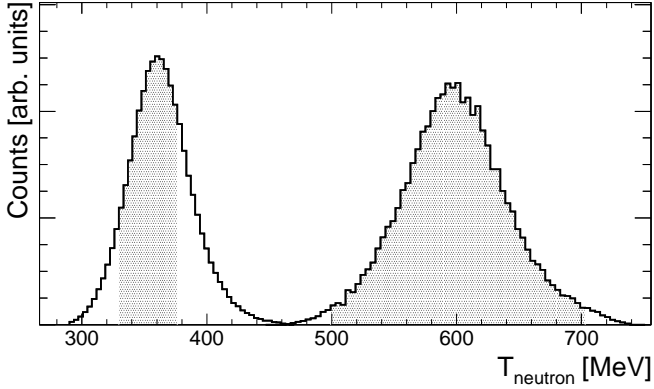}

\caption{\label{energies} Effective neutron beam energies for the quasi-free
$np\to dX$ reaction obtained with a 6~MeV spectator energy cut for those
events that fall within the $\pm2.5\sigma$ of the $\pi^0$ peaks in
Fig.~\ref{mm}. The results from the two experiments are shown on the same
plot and data within the shaded areas were retained in the analysis. These
correspond to $333<T_{\rm neutron} < 373$~MeV and $500<T_{\rm neutron} <
700$~MeV for the two beam energies.}
\end{center}
\end{figure}

The distributions in effective neutron beam energy for all the $dp\to p_{\rm
spec}dX$ events that fall within the $\pm2.5\sigma$ of the $\pi^0$ peaks in
Fig.~\ref{mm} are shown in Fig.~\ref{energies}. In the data analysis the
wings of the distributions were cut and only data in the regions $333<T_{\rm
neutron} < 373$~MeV and $500<T_{\rm neutron} < 700$~MeV were retained. In the
first case the average neutron energy varied between 351~MeV and 357~MeV,
being lowest for $\theta_{\pi}\approx 90^{\circ}$. The smaller angular range
at the higher energy resulted in much smaller deviations from the overall
average value of 600~MeV.

The analysing power in the \npdpi\ reaction~\cite{ARN1993} can be used to
estimate the average polarisation $Q$ of the hydrogen in the target cell. In
the 600~MeV experiment an unpolarised deuteron beam was part of the
three-polarisation-mode cycle~\cite{MCH2013} but only a small fraction of the
data were taken with unpolarised hydrogen in the cell. Using an extended data
set we find $Q^{\uparrow} = 0.59 \pm 0.07$ and $Q^{\downarrow} = -0.79 \pm
0.10$~\cite{MCH2013}, though it must be stressed that the uncertainty in
$Q^{\uparrow}- Q^{\downarrow}$, which does not depend on the unpolarised
target data, is much less than those of the individual modes. The analogous
figures at 353~MeV are $Q_\uparrow=0.59\pm 0.07$ and $Q_\downarrow=-0.70\pm
0.11$~\cite{DYM2013}. The consistency of these results suggests that the
polarisation of the gas in the target cell fed by the ABS is reproducible,
though one cannot rely on this when determining the spin-correlation in a
reaction.

The beam polarisations were also estimated using the asymmetry observed in
the $dp\to p_{\rm spec}d\pi^0$ reaction. However, it should be noted that the
selection of the polarisation modes for the deuteron beam was influenced by
the purposes of the principal experiments in the two cases. At 353~MeV, two
modes with nominal deuteron vector polarisations of $\pm\frac{2}{3}$ were
chosen with zero deuteron tensor polarisation. The measured neutron
polarisations in the deuteron beam were $P_\uparrow=0.55\pm 0.08$ and
$P_\downarrow=-0.45\pm 0.08$~\cite{DYM2013}.

In order to avoid complications arising from tensor polarisation components
of the deuteron beam, only the unpolarised and the nominal $(P_y,P_{yy}) =
(-\textstyle{\frac{2}{3}},0)$ modes were taken into account at 600~MeV and
the latter gave rise to a neutron polarisation that was measured to be
$P_\downarrow=-0.51\pm0.05$~\cite{MCH2013}.

The values obtained for the average of the product of the beam and target
polarisations using the above data do not have sufficient precision to
provide the best measurements of a spin correlation. Though both the beam and
target polarisations seemed stable to better than 10\%, we must also guard
against the possibility that these vary with time in different ways so that
the average of the product might differ from the product of the averages. In
both experiments one can use other reactions that are directly sensitive to
the average of $PQ$, where $Q\equiv |Q_\uparrow-Q_\downarrow|/2$, and
similarly for $P$.

By imposing the condition on the $\pol{n}\,\pol{p}\to \{pp\}_{\!s}\pi^-$
measurement that $\Ayy=1$ at 353~MeV, an average $PQ=0.373\pm 0.015$ was
obtained~\cite{DYM2013}. At 600~MeV per nucleon the theoretical
predictions~\cite{CAR1991} for the transverse spin correlation at small
momentum transfers in the $\pol{d}p\to \{pp\}_{\!s}n$ reaction yielded
$PQ=0.372\pm0.010$. The error bar here includes the uncertainty in the
correction coming from the difference between $|Q_\uparrow|$ and
$|Q_\downarrow|$ but it does not reflect the uncertainty in the reaction
model nor in the neutron-proton charge-exchange amplitudes~\cite{ARN2000}
used in the estimations. By looking at the quality of the predictions of
other charge-exchange observables at this energy~\cite{MCH2013}, it is
estimated these contribute to a systematic uncertainty in the value of $PQ$
of up to 10\%.

\begin{figure}[h]
\begin{center}
\includegraphics[width=1.0\columnwidth, angle=0]{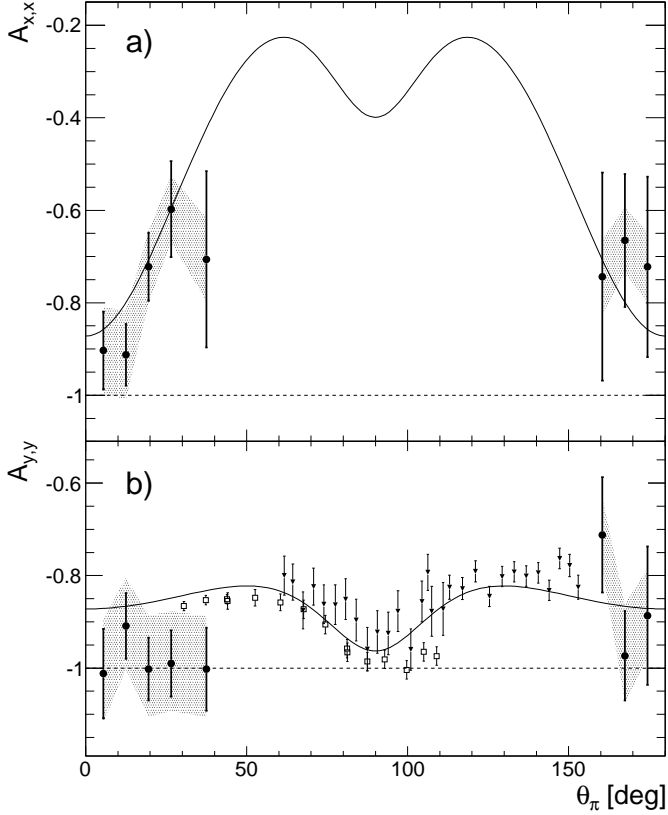}
\caption{\label{res600}The values of a) \Axx\ and b) \Ayy\ measured in the
\vnvpdpi\ at energies around 600~MeV as a function of the pion polar angle
$\theta_{\pi}$ in the c.m.\ frame. Statistical uncertainties are shown with
error bars; systematic uncertainties are illustrated with shaded bands. The
results are compared with the SAID predictions~\cite{ARN1993}, which have
been weighted with the measured energy distribution. Also shown are the PSI
\vpvpdpi\ results at 578~MeV (closed triangles)~\cite{APR1984} and those of
LAMPF at 593~MeV (open squares)~\cite{TIP1987}. The systematic uncertainties
in these cases are between 5\% and 10\%. }
\end{center}
\end{figure}

The procedures for extracting the values of the spin correlations are very
similar to those used to obtain \Axx\ and \Ayy\ for the quasi-free
$\pol{n}\pol{p}\to\{pp\}_{\!s}\pi^-$ reaction at 353~MeV~\cite{DYM2013} or
$\pol{d}\pol{p}\to \{pp\}_{\!s}n$ at 600~MeV per nucleon~\cite{MCH2013}. In
both cases the $\varphi_{\pi}$ dependence of Eq.~(\ref{asymmetry}) has to be
fitted in order to extract \Axx\ and \Ayy\ separately. At 600~MeV per
nucleon, the limited ANKE angular acceptance, especially in the vertical
direction, means that only the near-forward or backward regions were covered
and this problems becomes more severe for \Axx. The results in this case,
shown in Fig.~\ref{res600}, are in reasonable agreement with the predictions
of the SAID \ppdpi\ partial wave analysis~\cite{ARN1993} averaged over the
neutron energy distribution shown in Fig.~\ref{energies}\footnote{Due to the
nearly linear energy behaviour, the energy averaging is actually of little
importance here provided that the predictions are made at the mean beam
energy of 600~MeV.}. In addition to the statistical errors, which include
those arising from the background subtraction, the systematic uncertainties
are shown by shaded bands. The effects considered, which can vary
significantly in importance between 353 and 600~MeV, include uncertainties in
the relative luminosities, in the polarisation product $PQ$, in the influence
of the longitudinal polarisations arising from the Fermi motion, and that of
the finite range in $T_{\rm neutron}$. These have been compounded
quadratically. In contrast, no significant effects resulted from any
differences between the up and down polarisations nor from the resolution or
binning.

It should be noted that there are no measurements at all of \Axx\ for
\vpvpdpi\ in the vicinity of 600~MeV and those for \Ayy\ are generally at
larger angles~\cite{APR1984,TIP1987}, as shown in Fig.~\ref{res600}b. The
large negative values obtained in the forward direction are to be expected
because at 600~MeV there is sufficient energy to produce a $\Delta(1232)$
plus a nucleon at rest in the c.m.\ frame and such an intermediate state
would lead to $\Axx=-1$ at $\theta_{\pi}=0^{\circ}$.

\begin{figure}[h]
\begin{center}
\includegraphics[width=1.0\columnwidth]{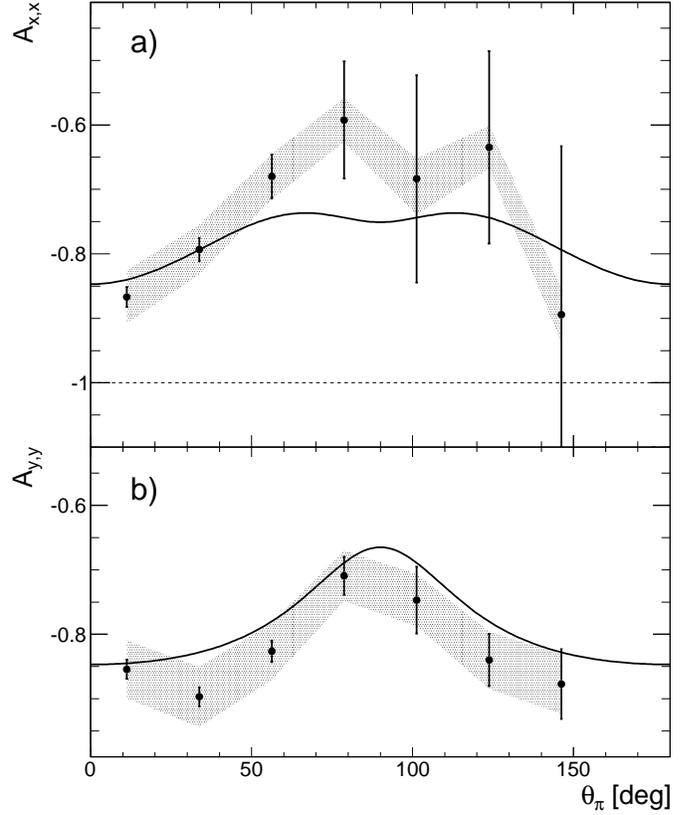}
\caption{\label{res353} Data obtained around 353~MeV, with the same
conventions as in Fig.~\ref{res600}. Statistical uncertainties are shown with
error bars; systematic uncertainties are illustrated with shaded bands.
Published data with very large error bars~\cite{PRZ2000} are not displayed.}
\end{center}
\end{figure}

The situation is qualitatively different for the 353~MeV data shown in
Fig.~\ref{res353} for there ANKE covers far more of the angular distribution.
It is also very different in that at such a low energy $s$-wave pion
production plays a more important role. The larger statistics allows a
tighter cut to be placed on $T_{\rm neutron}$ and our measurements of both
observables are well described by the current SAID partial wave
solution~\cite{ARN1993}. However, the existing data for \Axx\ and \Ayy\ for
the \vpvpdpi\ reaction at low energy have error bars as large as the
signal~\cite{PRZ2000} and thus provide little constraint on the partial wave
solution.

\begin{figure}[h]
\begin{center}
\includegraphics[width=1.0\columnwidth]{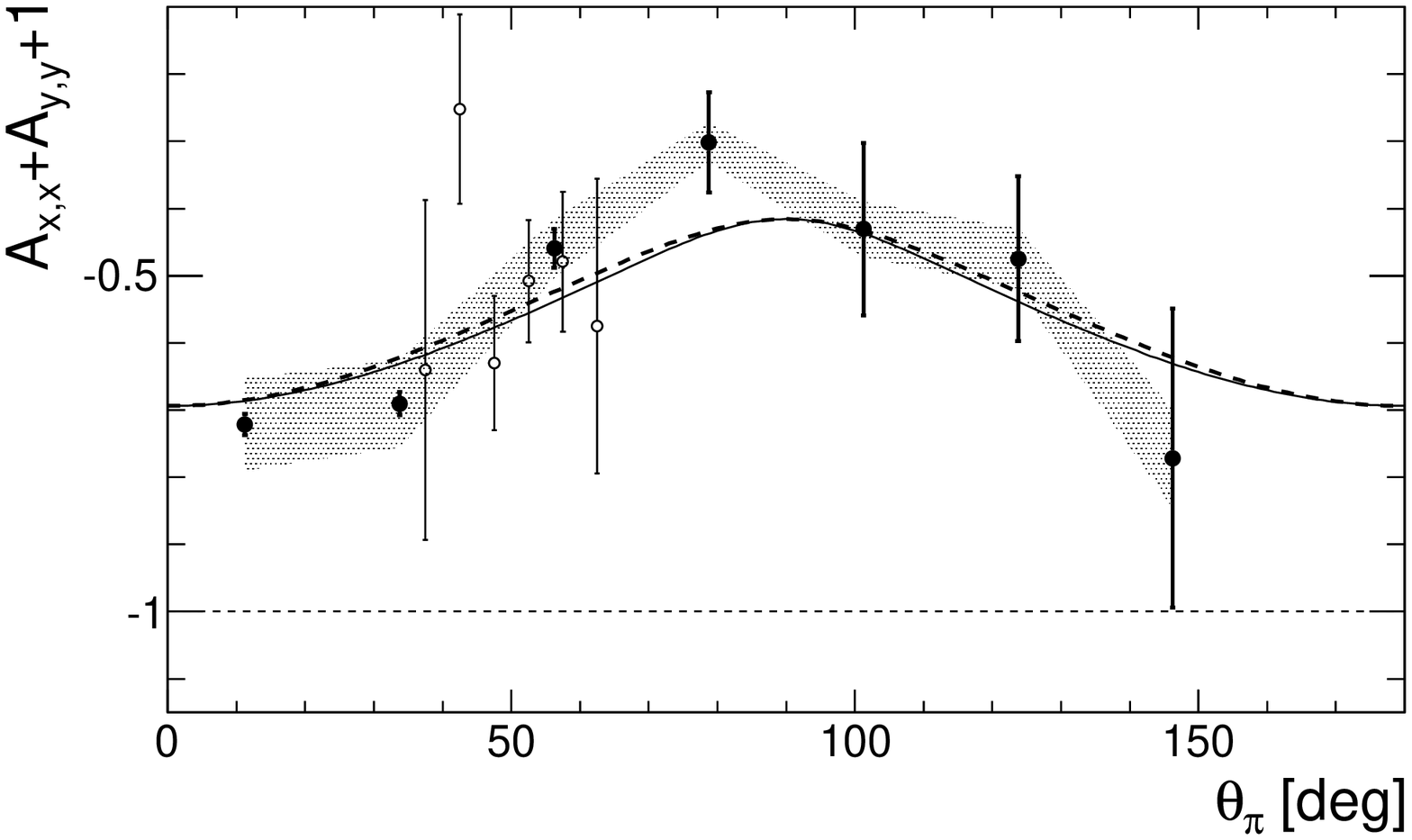}
\caption{\label{fig:Azz} The combination $1+\Axx+\Ayy$\ measured in the
\vnvpdpi\ reaction at 353~MeV as a function of $\theta_{\pi}$ compared with
the SAID \vpvpdpi\ predictions (dashed curve)~\cite{ARN1993}. Statistical
uncertainties are shown with error bars; systematic uncertainties are
illustrated with shaded bands.  Also presented are IUCF data for \Azz\ taken
at 350~MeV for the \vpvpdpi\ reaction (open circles)~\cite{PRZ2000} and the
SAID prediction for this observable (solid curve).}
\end{center}
\end{figure}

The measurements of the \vpvpdpi\ longitudinal spin correlation \Azz\ at
350~MeV are more precise~\cite{PRZ2000} and we compare these in
Fig.\ref{fig:Azz} with the values that we have obtained for $1+\Axx+\Ayy$\
for the \vnvpdpi\ reaction at 353~MeV. These two quantities coincide exactly
at $\theta_{\pi}=0^{\circ}$ and $90^{\circ}$ and they will be equal more
generally if one neglects pion $d$-waves, which should be a good
approximation at these low energies. The agreement with both these data and
the predictions of the SAID partial wave analysis is very reasonable.

We have presented here measurements of the \Axx\ and \Ayy\ spin correlation
coefficients for the \vnvpdpi\ reaction at two energies, one close to
threshold where the $s$-wave pion production is important and the other in a
region where $p$-wave production is dominant, being largely driven by the
$S$-wave $\Delta(1232)N$ intermediate state. All our data are consistent with
the current SAID solution for \ppdpi, though our measurements cover the small
angle region that is largely absent from existing \vpvpdpi\ data. No sign is
found for any breaking of isospin invariance.

The forward direction is in a region that is well suited for measurements
with ANKE and there is a particularly simple interpretation of the results
there. The value of \Axx\ at $\theta_{\pi}=0^{\circ}$ determines
unambiguously the ratio of the production of odd pion partial waves from an
initial spin-singlet $NN$ system to that for even partial waves that come
from spin-triplet initial states.

We are grateful to other members of the ANKE Collaboration for their help
with the experiment and to the COSY crew for providing such good working
conditions. This work has been supported by  the COSY FFE, the Shota
Rustaveli National Science Foundation, the Heisenberg-Landau programme, and
the European Union Seventh Framework Programme under grant agreement
n$^{\circ}$283286.

%
%

\end{document}